\documentclass[prd,aps,preprintnumbers,preprint,nofootinbib]{revtex4}
\usepackage{epsfig}
\usepackage{amsmath}
\usepackage{hyperref}

\def\bea {\begin{eqnarray}}
\def\eea {\end{eqnarray}}
\def\nn{\nonumber}

\def \ep {\epsilon}
\def \vep {\varepsilon}

\def \tp {{\tilde +}}
\def \tm {{\tilde -}}
\def \ca {{\cal A}}

\ifx\pdfoutput\undefined
\input epsf

\def\figscale#1#2{\epsfxsize=#2\epsfbox{#1.eps}}
\else

\def\figscale#1#2{\pdfximage width#2 {#1.pdf}\pdfrefximage\pdflastximage}
\fi

\begin{document}

\preprint{YITP-SB-13-4}

\renewcommand{\thefigure}{\arabic{figure}}

\title{Perturbation theory in (2,2) signature}

\author{Stanislav Srednyak and George Sterman}
\affiliation{C.N.\ Yang Institute for Theoretical Physics, Stony
Brook University, Stony Brook, New York 11794--3840, USA}

\date{\today}

\begin{abstract}
We identify a natural analytic continuation in four dimensions from Minkowski signature to a 
signature with two time-like momentum components.   
For two, three and four-point diagrams at fixed external momenta,
this continuation can be implemented as a countour deformation
that leaves dependence on the momenta unchanged.
   For arbitrary ultraviolet-finite scalar diagrams
it is possible to do two integrals
per loop in terms of simple poles in the new signature.    
This results in a representation of any such diagram as a sum of terms, each with two remaining integrals per loop.
\end{abstract}
\maketitle

\section{Introduction}

Scattering amplitudes are to a large extent determined by their singularity
structure in the complex planes of external momenta \cite{Eden:1966,Hwa:1966}.   
This feature has been exploited, for example,  
to derive a recursive construction for tree amplitudes \cite{Britto:2005fq}
from singularities at unphysical momenta, and enables the development of
unitarity-based techniques \cite{Bern:1994zx,Bern:1994cg} for the evaluation of loop integrals \cite{Britto:2010xq,Ellis:2011cr}.

For the construction of scattering amplitudes, any diagram in perturbation theory can be thought of as a  multidimensional 
complex integral, in the first instance by a Wick rotation from 
Euclidean space.  The rotation effectively changes a free Euclidean Green function,
$1/(- k_1^2 - \dots - k_n^2 - m^2)$ to the causal propagator, $1/(k_0^2 - k_1^2  \dots - k_{n-1}^2 - m^2+i\ep)$.
In this sense, the choice of contour corresponds to a change in the signature
of the metric, from all minus (or plus) to $(1,3)$.

Thus, the difference between Euclidean and Minkowski
Green functions can be thought of as a difference in the choice of contour integration.
 It is therefore natural to study
other signatures, corresponding to other choices of contour, in particular,
a $(2,2)$ signature, for which $k^2 = k_0^2 + k_1^2 - k_2^2 -k_3^2$.
In this connection, it is of interest to ask how to construct a perturbation theory  
based on this signature as an analytic continuation of Minkowski, and therefore ultimately
Euclidean, perturbation theory.  

The symmetries characteristic of $(2,2)$ signature 
help relate momentum to twistor spaces through a Fourier transform \cite{Witten:2003nn}.
The relationship between perturbation theory in $(2,2)$ 
and Minkowski formulations \cite{ArkaniHamed:2009si}, however, appears to be subtle and not yet fully clarified.  
Toward this goal, we will show below that there exists a non-singular
analytic continuation for scalar diagrams, analogous to Wick rotation, from Minkowski to $(2,2)$ signature  that crosses no 
singularities.   Perhaps surprisingly, singularities in the rotated integrals are avoided by the same
``$i\ep$" prescription as with Minkowski signature.   Theories with `two times' have also been
studied for their own interest \cite{Bars:2000qm,Bars:2008ru}, and most of our results below apply when
the number of spatial dimensions is greater than two.

In the process of the transformation from $(1,3)$ to $(2,2)$, both internal loop integration contours and external momenta are 
continued in terms of a single angular variable.    This naturally takes off shell any external momenta
that are on the light cone for Minkowski signature, except for momenta with no components in
the transverse direction that is rotated.   We observe that for such momenta overlapping collinear-infrared singularities survive the rotation,
and clarify a subtlety in the use of light cone coordinates that can lead to an apparent vanishing of
otherwise nonzero integrals.   More generally, for two, three and four-point functions, 
Lorentz invariance always allows us to choose momenta for which the Minkowski and $(2,2)$
functions are identical.    This result holds for massive and massless internal and external lines,
on shell and off-shell.

We begin the explicit development of these results in Sec.\ \ref{sec:construct},
where we show how to construct perturbation theory for $(2,2)$ signature
by a Wick-like rotation from Minkowski space, and discuss similarities and differences 
in the singularity structure of diagrams evaluated in $(1,3)$ and $(2,2)$ signature.   
In $(2,2)$ signature it is natural to introduce two sets of light cone coordinates, and in Sec.\ \ref{sec:lc-variables} 
we use this approach to show that after integration over the two ``minus" components
of each loop, the remaining $2L$-dimensional integrals are over a finite region, dependent
on the external momenta.   We also observe that in $(2,2)$ signature, perturbative unitarity
is realized in two different ways.    Restricting ourselves to ultraviolet finite diagrams, in Sec.\ \ref{sec:covariant-rep}
we derive a representation for an arbitrary $(2,2)$ scalar diagram as a $2L$-dimensional
integral.   We go on in Sec.\ \ref{sec:one-loop} to derive a compact representation for one-loop integrals
 with arbitrary masses and external momenta,
and illustrate how infrared singularities manifest themselves in $(2,2)$ signature,
using our representation for the box diagram.   We close with a summary of our results.

\section{From Minkowski to $(2,2)$}
\label{sec:construct}

As indicated above, our guiding criterion for the definition of  $(2,2)$ integrals is that they be analytic continuations
of corresponding integrals in Minkowski space, constructed so that the continuation manifestly
encounters no singularities.   In fact, such a construction can be carried out by a direct 
generalization of Wick rotation.   
In this discussion, we restrict ourselves to scalar integrals only.     
Like Wick rotation, the construction turns out to be completely general and rather simple.
We give it below, followed by a few consequences.

\subsection{Defining the integrals}

We consider an arbitrary perturbative
integral, written in covariant form, with $L$ loops and $N$ lines of arbitrary mass, possibly with positive imaginary parts,
and with external momenta $p_j$,
which may or may not be on shell,
\bea
I_{N,L}(p_j)
= 
(-i)\, i^{L-1}
\prod_{{\rm loops}\ a=1}^L
\int \frac{d^4 l_a}{(2\pi)^4}\
\prod_{{\rm lines}\ i=1}^N \
\frac{1}{k_i^2(l_a,p_j)-m_i^2+i\ep}\, .
\label{eq:I_LN}
\eea
We take $k^2 = k_0^2 - k_1^2-k_2^2-k_3^2$ to start.   The first factor of $-i$ on the right hand side normalizes 
tree diagrams to be real whenever each vertex is associated with a factor $-i$ and each line with an $i$.   Here
and below, we set the coupling constant to unity.
As indicated in Eq.\ (\ref{eq:I_LN}), line momenta are themselves determined by the loop and external momenta,
through linear combinations that can by summarized by matrices $\eta_{ia}$ and $\xi_{ij}$, respectively,
\bea
k_i = \eta_{ia} l_a + \xi_{ij}p_j\, ,
\label{eq:ki-sum}
\eea
with $\eta_{ia},\, \xi_{ij}=\pm1,\, 0$.
The integration contours are defined, as usual, by the ``$i\ep$" prescription, in which energy
integrals pass above the pole at the larger on shell energy for each propagator, and below the pole at the smaller on shell energy.    

\begin{figure}
\centerline{\figscale{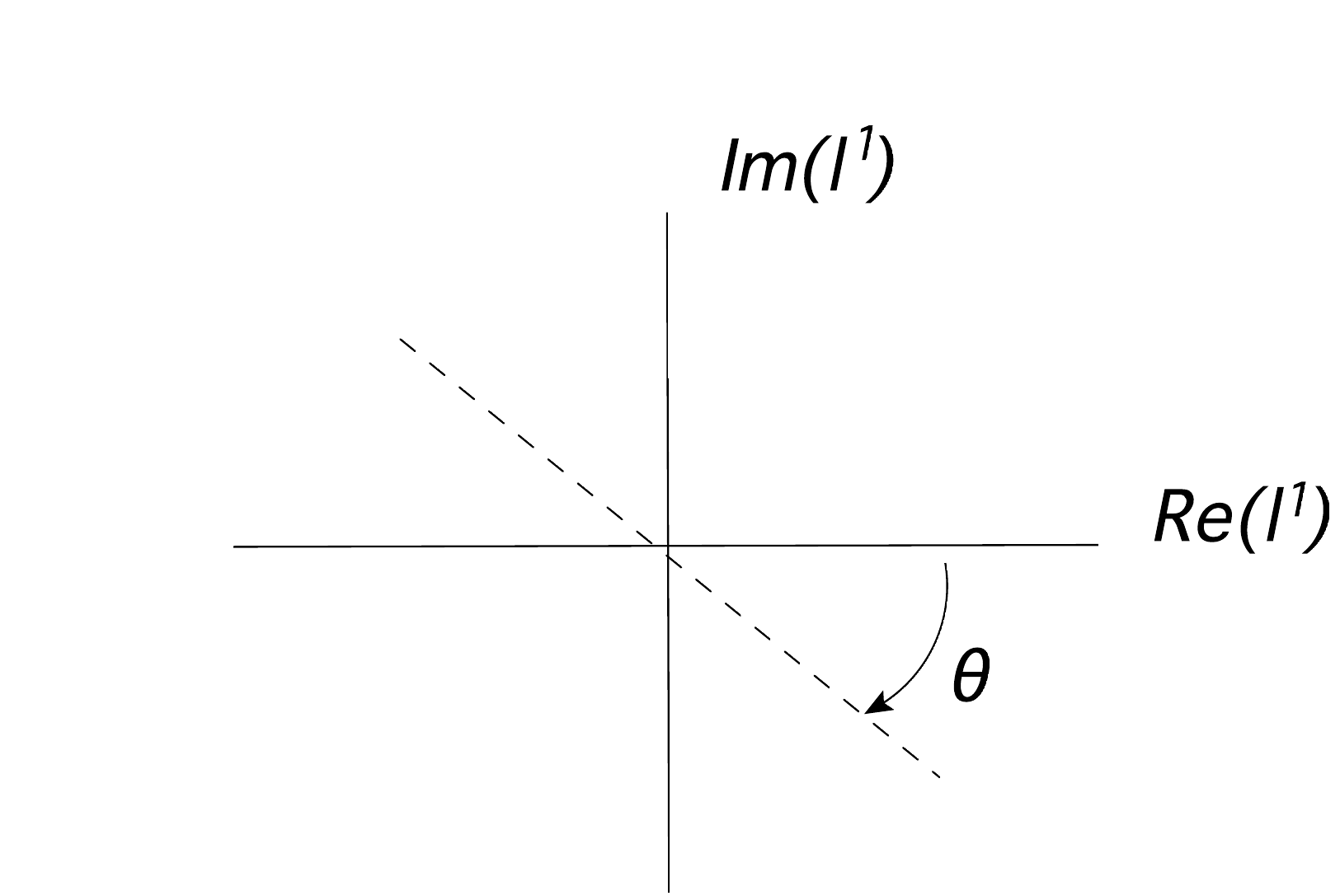}{9cm}}
\caption{Rotation of the $l^1$ contour.}
\label{fig:rotate}
\end{figure}

We now define a new parameter, $\theta$, and a new function, $I_{N,L}(p_j,\theta)$, constructed so
that it equals the original diagrammatic integral, (\ref{eq:I_LN}) at $\theta=0$,
\bea
I_{N,L}(p_j,\theta=0) = I_{N,L}(p_j)\, .
\label{eq:I-boundary}
\eea
The new function is defined in terms of momentum components, as a joint rotation of the `one' components,
 $p_j^1$ of all external and $l_a^1$ of all loop momenta, as illustrated in Fig.\ \ref{fig:rotate},
\bea
I_{N,L}(p_j,\theta)
&=& (-i)\, i^{L-1}
\prod_{{\rm loops}\ a =1}^L
\int \frac{d l^0_adl^3_a dl_a^2 }{(2\pi)^4}\ \int_{-\infty}^\infty d(l_a^1\, e^{-i\theta})
\nn\\
&\ & \hspace{-25mm}\times
\prod_{{\rm lines}\ i} \
\frac{1}{\left(k_i^0(l^0_a,p^0_j)\right)^2 - \left(\eta_{ia} l_a^1e^{-i\theta} + \xi_{ij}p_j^1e^{-i\theta}\right)^2 - \left(k_i^2(l_a^2,p_j^2)\right)^2
-\left(k_i^3(l^3_a,p^3_j)\right)^2 -m_i^2+i\ep}\, .
\nn\\
\label{eq:I_LN_rotate}
\eea
At finite $\theta$, the real and imaginary parts of the denominator of propagator $i$ are given by
\bea
{\rm Re}(k_i^2+i\ep) &=& (k_i^0)^2 - \left(\eta_{ia} l_a^1 + \xi_{ij}p_j^1 \right)^2\, \cos(2\theta) - (k_i^2)^2 -(k_i^3)^2 
\nn\\
{\rm Im}(k_i^2+i\ep) &=&  \left(\eta_{ia} l_a^1 + \xi_{ij}p_j^1\right)^2\, \sin(2\theta) \ +\ \ep\, .
\label{eq:re_im}
\eea
As we vary  $\theta$ from zero to $\frac{\pi}{2}$, the coefficient of the square of $(k_i^1)^2$ in the real part
changes sign, while the imaginary part of each diagram starts at $+i\ep$, increases to a 
maximum at $\theta=\frac{\pi}{4}$, always staying positive, and decreases back down to
$+i\ep$ at $\theta = \frac{\pi}{2}$.   For fixed values of the original momenta, $p_j$, the integrand
is thus finite over the entire continuation in $\theta$, and crosses no singularities.    
The procedure works for any choice of masses, so long as their imaginary parts are
positive.

The result of this procedure, continuation from $\theta=0$ to $\theta=\frac{\pi}{2}$, is a
smooth transition from Minkowski signature, with a single time-like momentum
component, to a $(2,2)$ integral, in which the $1$ component has joined the $0$ component
as a positive contribution to the invariant squares of the momenta.    This fully-rotated integral is 
given explicitly by
\bea
I_{N,L} \left (p_j,\frac{\pi}{2} \right)
&=& -\
\prod_{{\rm loops}\ a =1}^{L+1}
\int \frac{d l^0_a dl_a^1\, dl^2_a dl_a^3 }{(2\pi)^4}
\prod_{{\rm lines}\ i=1}^N \
\frac{1}{\left(k_i^0\right)^2 + \left(k_i^1\right)^2 -\left(k_i^2 \right)^2  - \left(k_i^3\right)^2 -m_i^2+i\ep}\, ,
\nn\\
\label{eq:I_LN_22}
\eea
where we have suppressed the linear dependence of line momenta on loops and external lines.
We note that the integrals are defined by the same $i\ep$-prescription as in Minkowski space,
a perhaps surprising result.    This definition has (at least) two important consequences for
the singularity structure of $(2,2)$ diagrams, which we develop in the following two subsections.

\subsection{Signature-invariance of two, three and four-point functions}

For two- three- and four-point functions in Minkowski space,
we can always go to a frame where at least one component of spatial momentum is zero for all external lines.   
For $2 \rightarrow 2$ scattering, for example, this is the normal to the scattering plane.   
If we choose this direction as the `one' direction above, all $p_j^1=0$, and the 
rotations of loop momenta can be carried out for fixed (Minkowskian) external momenta
without crossing singularities.   Indeed, if the momentum integrals are convergent, 
Cauchy's theorem ensures that the integrals are independent of $\theta$, because the
rotation can be treated as the change of a contour that can be closed at infinity.    As a result,
for such diagrams, we have
\bea
A_n^{(3,1)}(p_1 \dots p_n) = A_n^{(2,2)}(p_1 \dots p_n)\, ,
\label{eq:13is22}
\eea
for $n\le 4$, so long as the extra time-like coordinate is chosen perpendicular to the space
spanned by the $p_i$, which remain in a Minkowskian $(1,2)$ subspace.    
Such a choice is always possible for $n\le 4$.   This result 
applies to scalar diagrams of all orders, any choices of (real) 
masses, and for off-shell Green functions as well as on shell amplitudes.
Indeed it applies to diagrams with any number of external lines so long as all $p_j^1=0$.   
We note that an analogous invariance applies to Wick rotation for diagrams with all $p_j^0=0$.

Although a simple consequence of analytic continuation, the relation (\ref{eq:13is22}) will enable us to give
new representations for loop integrals in $(1,3)$ signature for two-, three- and four-point functions 
in Minkowski space, as special cases of general representations of $n$-point functions
in $(2,2)$.    These representations will follow from the introduction of double light cone coordinates
in $(2,2)$ signature, which we will describe in Sec.\ \ref{sec:lc-variables}.    We turn first, however, to a brief
investigation of the singularity structure of $(2,2)$ integrals.

\subsection{Singularities  in $(2,2)$}
\label{subsec:singularities}

Starting from the defining equation (\ref{eq:I_LN_22}), we can make quite strong statements
about the origin of the singularities of perturbative integrals in  $(2,2)$ signature.
In particular, because they share the same $i\ep$ prescription with their $(1,3)$ counterparts,  the Landau equations 
\cite{Eden:1966,Landau:1959fi,Bjorken:1965zz} that help determine singularities in perturbative integrals take the same form
for $(1,3)$ and $(2,2)$ signatures, Eqs.\ (\ref{eq:I_LN}) and (\ref{eq:I_LN_22}).   
This is most easily confirmed by reviewing the use of Feynman parameterization to identify possible
pinches in loop integrals \cite{Eden:1966}, to emphasize its signature independence.     For the $(2,2)$ case, for example, we have simply
\bea
I_{N,L}\left(p_j,\frac{\pi}{2} \right)
&=& -\
\Gamma(N-1)\
\prod_{{\rm loops}\ a =1}^L
\int \frac{d l^0_a dl_a^1\, dl^2_a dl_a^3 }{(2\pi)^4}
\ \prod_{{\rm lines}\ i=1}^N \  \int_0^1\, d\alpha_i\ \delta\left( 1 - \sum_{i=1}^N \alpha_i\right) 
\nn \\
&\ & \hspace{10mm} \times\ 
\frac{1}{\left [ \sum_{i'=1}^N \alpha_{i'} \left[ k_{i'}^2 (l_a,p_j) - m_{i'}^2 \right] +i\ep \right]^N }\, ,
\nn\\
\label{eq:I_LN_Feynparam}
\eea
the difference from $(1,3)$ being entirely in the definition of the $k_i^2$ on the
right hand side, and the argument on the left.
Because line momenta $k_{i'}$ are linear in loop momenta $l_a$, the single,
parameterized denominator is quadratic in every loop momentum component 
$l_a^\nu$, while being linear in the parameters $\alpha_{i'}$.   
We note that the relative signs of the denominator terms in this
expression are determined uniquely by requiring that the coefficient of the 
imaginary term $i\ep$ be $\alpha_i$-independent.    This ensures that whatever
component integral we do first has one $N$th order pole in the upper half plane, and one in the lower half plane.

Necessary conditions for the presence of a  singularity in (\ref{eq:I_LN_Feynparam}) are then that those 
line momenta $k_{i'}$ whose
coefficients $\alpha_{i'}$ are nonzero must satisfy 
\bea
\frac{\partial }{\partial l_a^\nu}\ \left [ \sum_{i'=1}^N \alpha_{i'} \left( k_{i'} (l_a,p_j) \right)^2 +i\ep \right] 
= 2\sum_{i'=1}^N \alpha_{i'} \eta_{a i'} k_{i'}^\nu (l_a,p_j) = 0\, ,
\label{eq:Landau}
\eea
for every component $\nu$ of every loop $l_a$, with $\eta_{a i}$ the matrix that relates loop to line momenta in 
Eq.\ (\ref{eq:ki-sum}) above.   These are the same (Landau) equations, whether 
in $(1,3)$ or $(2,2)$.
 A singularity also requires, of course, that $k_i^2=m_i^2$
for the relevant lines.   Thus, given the differences in the signatures that define $k_i^2$ for $(1,3)$ and $(2,2)$,
there is no immediate correspondence between momentum configurations found in the
two cases for the same diagrams.   In particular, it is not obvious whether there is an analog  in $(2,2)$ 
of the Coleman-Norton criterion for singularities \cite{Coleman:1965xm} in $(1,3)$, that on shell momenta
at a singularity correspond to 
a physical scattering process.   This would at least require us to develop 
intuition on what  ``physical scattering" means in $(2,2)$ signature.
Nevertheless, although we do not have such a general criterion for singularities in $(2,2)$, we can make some significant
observations, finding a wide range of both similarities and differences from $(1,3)$.   

In this connection, we note a simple result on singularity surfaces for Green function integrals like $I_{N,L}(p_j,\theta)$, Eq.\ (\ref{eq:I_LN_Feynparam}).
When the external lines of a diagram are restricted to a subspace where one component vanishes for all lines,
\bea
p_j^\nu=0\, , \quad {\rm all}\ j\, ,
\label{pjnu-zero}
\eea
the corresponding component of all {\it internal} on-shell  lines must vanish at any pinch singularity.
To see this, consider an arbitrary  ``candidate" pinch surface with a set of on shell lines, $k_l$, $k_l^2=m_l^2$,
some of which have nonzero component $k_l^\nu$.
Starting with any line momentum $k_i \in \{k_l\}$ with $k_i^\nu\ne 0$, we can follow the flow of positive (or negative) $k_i^\nu$,  
from line $i$ into some unique vertex of the diagram, which we label as, say, $v_0$.  Let us consider the combination $k_i,\, v_0$ as the
beginning of a path (a ``chain") through the diagram.  We continue the path by picking any line attached
to vertex $v_0$ that carries positive $\nu$ component out of $v_0$ to some other vertex $v_1$.   
By momentum conservation, there must be at least one such line.   In this way, we 
continue the path through the diagram.   Because of our assumption (\ref{pjnu-zero}),
the  $\nu$ component can never flow out of the diagram, and therefore the path 
will stay inside the diagram at each step.   If the diagram is of finite order, the path 
will eventually intersect itself, by connecting a sequence of vertices,
\bea
v_0 \rightarrow v_1 \rightarrow  \cdots \rightarrow v_n \rightarrow v_0\, .
\label{eq:nu-loop}
\eea
In general, there is more than one such path if the diagram has more than one loop,
but in any case we can pick a loop momentum $l_a$ that flows precisely around the loop
specified by the sequence of vertices (\ref{eq:nu-loop}).   For this loop, all the factors $\eta_{ia}$ and all the
$\nu$ components of lines $k_i$ are positive, and the Landau equations
 (\ref{eq:Landau}) cannot be satisfied for nonzero $\alpha_i$.   Therefore, this set of 
 lines, and since they are arbitrary any set of lines with nonzero $k_i^\nu$, cannot satisfy the Landau equations and 
 cannot be pinched on shell.
 
 This result shows us that 
a kinematic range where the two signatures give a similar singularity structure can be found for $2\rightarrow 2$ on shell  scattering
amplitudes,
\bea
p_1 + p_2 \rightarrow p_3+p_4\, , \quad p_i^1\ =\ 0\, , \quad p_i^0\ >\ 0\, .
\eea
For such a process, no pinch surface can have internal lines with a one component, and the classification
of pinch singularities follows the same reasoning as in Minkowski space \cite{Akhoury,Sen83,Botts:1989kf}.   
It is worth pointing out that in Minkowski space, because the scattering is planar in the center-of-mass,
pinch surfaces are always restricted to a three-dimensional subspace here as well.   
Recall that we have observed above that the continuation can be carried out without changing external momenta in this frame.
The only difference in $(2,2)$ compared to $(1,3)$ is that the ``normal" to this subspace is now a time-like rather than a space-like variable.
In particular, for fixed angle scattering in massless theories \cite{Akhoury,Sen83,Botts:1989kf}, pinch surfaces in $(2,2)$ reduce to the same  ``jet", ``soft" and ``hard" subdiagrams
long known to characterize these amplitudes in Minkowski space.
We will not pursue a further investigation of this case here, but only note that there is every reason to believe that for gauge theories the
basic factorization properties and infrared structure of massless Minkowski $2\rightarrow 2$ amplitudes \cite{Sen83,Botts:1989kf} are the same in $(2,2)$.

 The fundamental similarity between $(1,3)$ and $(2,2)$ singularity structure for $2\rightarrow 2$ amplitudes is
 certainly the exception, and 
 we need not look far for fundamental differences, once we relax the condition $p_j^1=0$,
 for external lines.    
 Indeed,  once the number of external lines exceeds four, this condition restricts us to a subspace
of their full momentum space.    In the new signature, a general amplitude has many
singularities that are qualitatively different from those found in Minkowski signature.

\begin{figure}
\centerline{\figscale{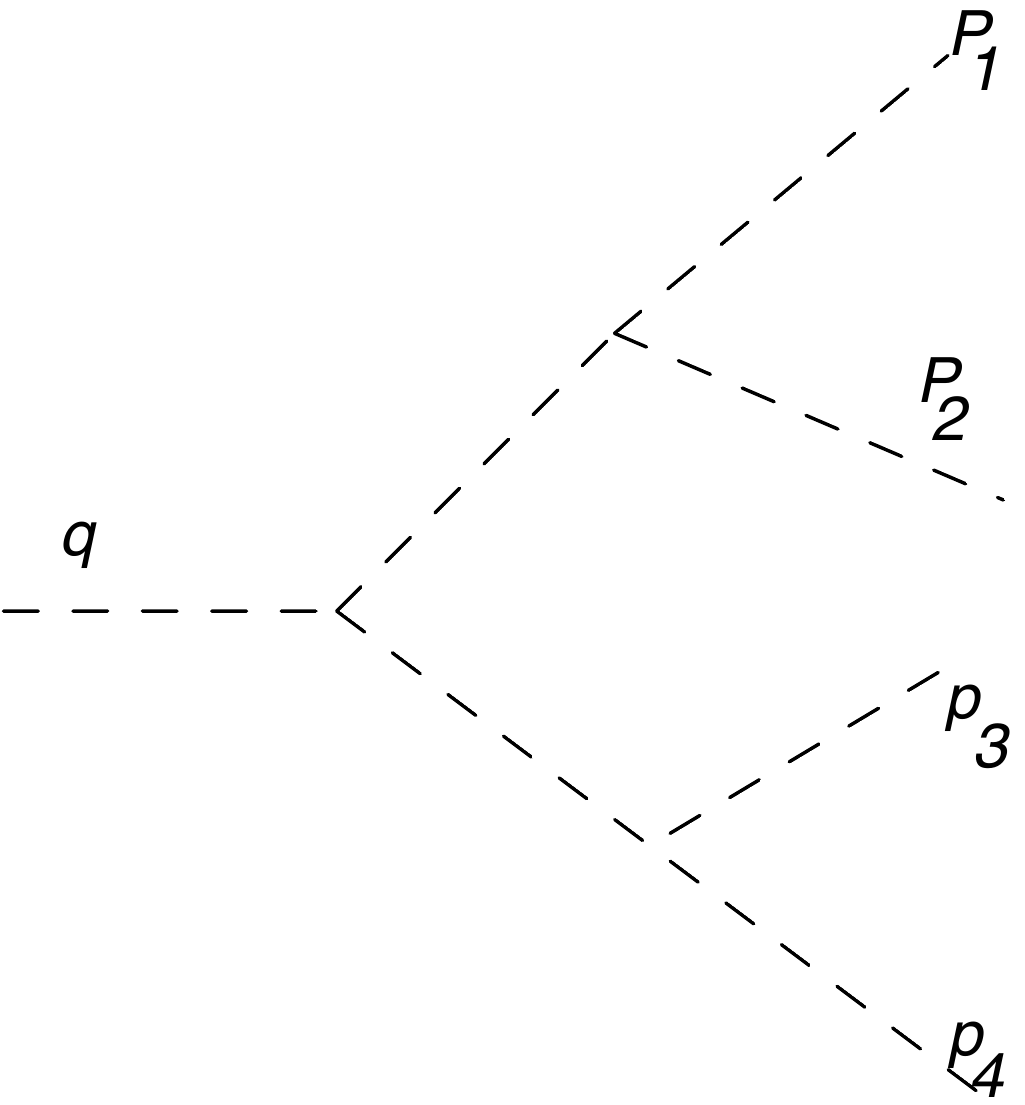}{5cm}}
\caption{One-to-four scalar process discussed in the text.}
\label{fig:4-jet22}
\end{figure}

A fundamental property
of light-like lines in Minkowski space is that the sum of two positive energy light-like momenta has a positive semi-definite
invariant mass, which vanishes only when the momenta are proportional, that is to say, the lines are collinear.
For $(2,2)$ signature, in contrast, every light-like momentum, $v^\mu,\, v^0>0$ defines
a one dimensional subspace of light-like vectors $\bar v^\mu$ with  $\bar v^2= \bar v\cdot v = 0$,
found by making equal $SO(2)$ rotations on the pairs $(v^0,v^1)$ and $(v^2,v^3)$, 
\bea
\bar v^\mu &=&  \left( \begin{array}{cc} R& 0 \\ 0 & R \end{array}\right)
\left (  \begin{array}{c} v^0 \\ v^1 \\ v^2 \\ v^3 \\ \end{array}\right)\, , \ R \in SO(2)\, .
\eea
As a result, in $(2,2)$, the sum of two, non-collinear light-like momenta can also be light-like.
This has consequences for the singularity structure even of tree diagrams, as illustrated by Fig.\ \ref{fig:4-jet22}.
Here we start with the generalized ``rest" momentum, $q^\mu= (Q,Q', 0, 0)$ in $(2,2)$ signature, and
we show a lowest-order diagram that produces four lines of momenta
\bea
p_1 &=& \left(\, \frac{Q}{2}\, , 0\, , 0\, , \frac{Q}{2}\, \right) \, ,
\nn\\
p_2 &=& \left(\, 0\, , \frac{Q'}{2}\, , \frac{Q'}{2}\, , 0 \, \right) \, ,
\nn\\
p_3 &=& \left(\, \frac{Q}{2}\, , 0\, ,0\, , -\ \frac{Q}{2}\, \right) \, ,
\nn\\
p_4 &=& \left(\, 0\, , \frac{Q'}{2}\, ,-\ \frac{Q'}{2}\, , 0\, \right) \, .
\eea
For this set of ``outgoing particles", the virtual lines have $(p_1+p_2)^2=(p_3+p_4)^2=0$,
in sharp contrast to the corresponding diagrams of Minkowski space whenever
the outgoing lines are noncollinear.    This suggests that beyond the simplest
amplitudes, the concept of ``jets", for example, would have to be generalized
in any complete picture of $(2,2)$ scattering.    

\bigskip

\section{Light Cone Variables}
\label{sec:lc-variables}

We now turn to another interesting feature of $(2,2)$ integrals that are ultraviolet convergent.
In the rotated integral, $I_{N,L}(p_j,\frac{\pi}{2})$, Eq.\ (\ref{eq:I_LN_22}), 
there is a nice symmetry between the pairs of components, $0,3$ and $1,2$,
and it is natural to introduce two pairs of light cone loop momentum variables,
\bea
k^{\tilde \pm} &\equiv&  k^1 \pm k^2 \, , \nn\\
k^\pm &=& k^0 \pm k^3 \, ,
\label{eq:kdefs}
\eea
where we have chosen a normalization for which
\bea
k^2 &=& k^+k^- + k^{\tilde +}k^{\tilde -}\, , 
\nn\\
2 k\cdot k' &=& k^+k'{}^- + k^- k'{}^+ + k^\tp k'{}^\tm + k^\tm k'{}^\tp\, ,
\nn\\
d^4 k &=& \frac{1}{4} dk^+ dk^{\tilde +} dk^- dk^{\tilde -}\, .
\label{eq:lc-norm}
\eea
We use these variables
below to develop a procedure for doing $2L$ integrals in an arbitrary
ultraviolet finite $L$-loop diagram.    Before doing so, we point out one subtle point 
in making such a change of variables.    This observation applies as well to the 
use of light cone variables in $(1,3)$ to develop, for example, light cone ordered perturbation theory \cite{Chang:1968bh}.

\subsection{Convergence and light cone variables}

Consider the manifestly finite two-dimensional integral, of a self-energy form,
\bea
I_2\left(p,m^2\right) &=&-i\ \int_{-\infty}^\infty \frac{dk_1 dk_0}{(2\pi)^2} \ \frac{1}{\left[ (k_0+p)^2- (k_1-p)^2 - m^2+i\ep \right]}\,
\frac{1}{\left[ k_0^2- k_1^2 - m^2+i\ep \right] }
\nn\\
&=& \frac{1}{4\pi\, m^2}
\, .
\label{eq:I2k1k2}
\eea
Here the two-dimensional ``external" momentum is $P=(P_0,P_1)=(p,-p)$, with $p>0$. 
The result of this integral is independent of parameter $p$ because $P^2=0$, and readily
follows from standard formulas based on Feynman parameterization and Wick rotation.
We can also evaluate (\ref{eq:I2k1k2})  as a pair of
 complex integrals explicitly in terms of its poles.    Each of the two variables, $k_0$ and $k_1$ encounters four poles, two  in each half plane,
and we can  perform the integral by closing one contour in either the upper or lower half plane without
Wick rotation.
   
Now let us try to re-express the integral, Eq.\ (\ref{eq:I2k1k2}) in terms of light cone coordinates, $k^\pm \equiv k_0 \pm k_1$, 
as in Eq.\ (\ref{eq:kdefs}).   In this notation, the vector $P$ has  $P^-=2p$ and $P^+=0$.  This, however, gives
\bea
I_2\left(m^2\right) = 
\frac{-i}{2}\, \int_{-\infty}^\infty \frac{dk^+ dk^-}{(2\pi)^2} \ \frac{1}{\left[ k^+(k^-+2p) - m^2+i\ep \right]}\
\frac{1}{\left [ k^+k^- - m^2+i\ep \right]}\, ,
\label{eq:I2kpkm}
\eea
which vanishes because the two poles in the $k^-$ integral are always on the same side of the contour,
regardless of the value of $k^+$.   This would seem to imply that the self energy vanishes whenever $P^+=0$,
a paradoxical result that would extend to four dimensions.   On the other hand, if we do the $k^+$ integral first,
the result is nonzero, because the two poles in $k^+$ are on opposite sides of the contour for $- P^- < k^-<0$.

 The reason for this inconsistency is that 
the change from Cartesian to light cone variables involves an exchange of integrals that are not uniformly
convergent in this case.   To be specific, suppose we wish to do the $k^-$ integral first at fixed $k^+$.   We would
then first change variables from (say) $k_0$ to $k^+$ in the original $k_0,k_1$ form, Eq.\ (\ref{eq:I2k1k2}) at fixed $k_1$, giving
\bea
I_2\left(m^2\right) &=& 
\frac{-i}{(2\pi)^2}\,
\int_{-\infty}^\infty dk_1 \int_{-\infty}^\infty dk^+ \ 
\frac{1}{\left [ (k^+-k_1+p)^2- (k_1-p)^2 - m^2+i\ep \right]}\, 
\nn\\
&\ & \hspace{35mm} \times\
\frac{1}{ [(k^+-k_1)^2 - k_1^2+i\ep] }
\,.
\label{eq:I2kpk1}
\eea
The next step would be to exchange the $k_1$ and $k^+$ integrals, and then change variables
from $k_1$ to $k^-$ at fixed $k^+$, giving (\ref{eq:I2kpkm}), but this is not possible 
because the unbounded $k_1$ integral diverges badly for $k^+=0$.   
We may note, however, that this pitfall does not prevent us from carrying out
rotations in Cartesian coordinates from $(1,3)$ to
$(2,2)$ as above.    The transition to light cone coordinates is a separate issue.

\subsection{Finite volume}
\label{sec:finite}

Having pointed out a subtlety associated with the vanishing of external
plus momenta, we can limit ourselves to all nonzero external plus momenta.
 In this case, we can do all the minus loop integrals in a given
diagram, to get a sum of terms given by the rules of light cone ordered perturbation theory (LCOPT) \cite{Chang:1968bh}. 
This procedure does not depend at all on whether or not we have carried out the rotation
that takes us from $(1,3)$ to $(2,2)$ signature.
   For a scalar diagram ${\cal G}$ (normalized as above so that  tree graphs are real) the LCOPT expression
 found by
integration over minus momenta is related to the covariant form  by 
\bea
{\cal G}(\{p_a\})
&\equiv& (-i)\, i^{L-1}\, 
\sum_{{\rm orderings}\ T}\ \int \prod_{{\rm loops}\,\{l\}} \frac{d^4 l}{4(2\pi)^4}\ \prod_{{\rm lines}\, k} \; \frac{1}{k^2-m_k^2+i\ep}
 \nn\\
 &=&
-\, \int \prod_{{\rm loops}\,\{l\}} \frac{dl^\tp dl^\tm dl^+}{4(2\pi)^3}\ \prod_{{\rm lines}\,\{k\}}\ \frac{\theta(k^+)}{k^+}\
\prod_{{\rm states}\,\{i\}\ {\rm in}\ T} \frac{1}{P_i^- - s_i\left ( [k]\right)+i\epsilon }\ \, ,
\nn\\
\label{eq:lcopt}
\eea
where $P_i^-=\sum_{a\in i} p_a^-$ is the algebraic sum of total incoming and outgoing minus momenta up to state $i$, and where
\bea
s_i\left ( [k]\right)
&=&
\sum_{{\rm lines}\,\{k\}\, \in\, {\rm state}\, i} [k]^-
\nn\\
&=& \sum_{k \in i} \frac{-k^\tp k^\tm + m_k^2 }{k^+}
\nn\\ 
&\equiv& \sum_{k \in i} \left(\, -\ k^{\tm} r_k + \mu_k\, \right) \, ,
\label{eq:mdef}
\eea
is the sum of all the on shell minus momenta in a specific state.   
We have written the result in terms of the $(2,2)$ 
signature transverse `light cone' variables formed from $k_T=(k^1,k^2)$ in Eq.\ (\ref{eq:kdefs}),
and we define
\bea
r_k &\equiv& \frac{k^\tp}{k^+}
\nn\\
\mu_k &\equiv& \frac{m_k^2}{k^+}\, ,
\label{eq:mudefs}
\eea
where the label $k$ identifies both the line momentum and the
corresponding mass.  
The transition to $(2,2)$ signature can be carried out before the minus
integrals that lead to the second equality in Eq.\ (\ref{eq:lcopt}), or after.   

We will first use the invariant integral representation of 
an arbitrary ultraviolet finite diagram in Eq.\ (\ref{eq:lcopt}) to show
that the volume of the $l^+$ integrals  is finite
after the $l^-$ integrals at fixed $l^\tp$ and $l^\tm$.  
We will go on to use the light cone ordered form to show that the $l^\tp$ 
integrals also have a finite volume after the integration
over the $l^\tm$ for diagrams that are ultraviolet finite.  

Assume, then, that some plus loop momentum
grows without bound in such a way that it is much larger than the 
corresponding components of all external momenta.    
As we shall see, it is then possible to find a minus loop integral
such that all of its poles are in the same half-plane, either upper or
lower.  Such an integral gives zero, and  because we assume that the diagrams are well-behaved at
infinity, we can choose to do this minus integral first.    We conclude that the
integral is non-zero only in a bounded region in plus momentum.
To be specific, let us provide an explicit construction of the loop in question,
by an argument similar to that of Sec.\ \ref{subsec:singularities} above.

The construction begins by identifying the internal line with the largest plus momentum, which we may call $K_1^+>0$.
We can choose the orientation of momentum flow so that this quantity is positive.
Momentum $K_1^+$ then flows into a unique vertex of the diagram, which we may call
$V_1$, and out of a unique vertex $V_0$.  Suppose that vertex $V_1$
is an $a$-point vertex.  Since momentum $K_1^+$ flows in to $V_1$ at least
one line must carry a momentum $K_2^+\ge K_1^+/(a-1)$ out of $V_1$.
 If $K_1^+$ is sufficiently large, this line cannot flow out of the diagram,
 but must flow to another vertex, $V_2$, internal to the diagram. 
 Assuming for simplicity that this is also an $a$-point vertex, at least one
 line must carry plus momentum $K_3^+\ge K_1^+/(a-1)^2$ out of $V_2$.
 We repeat the process, following the largest flow of plus momentum,
 and in each case, we find a momentum that flows out of the next
 vertex that is proportional to $K_1^+$, and which therefore cannot
 carry momentum onto an external line when $K_1^+$ is large enough.
For any diagram of finite order, we will eventually encounter a vertex
$V_m = V_k$, with $k=0 \dots m-2$ ($m=2$ is not possible for a diagram
with no ultraviolet-divergent subdiagrams in four dimensions).    This is the loop we are after.

Exactly the same reasoning would apply to show that the $l^\tp$ integrals also have a finite
volume at fixed $l^+$ and $l^-$.   
We show next, however, that the $l^\tp$ integration regions are limited even {\it after} the $l^-$ integrals
are performed.   For this, we apply a similar reasoning to the light cone ordered expression, the second equality in Eq.\ (\ref{eq:lcopt}).   
That is, we assume that some set of  loop momenta, $\{l_a^\tp\}$ become large enough that it is possible
to find a loop around which every line carries plus tilde momentum in the direction of the loop.
We claim that in this case, the momentum $l_b^\tm$ that flows around this loop sees poles only
in the lower (or upper) half plane in Eq.\ (\ref{eq:lcopt}), so that its integral vanishes.   To show this, we
consider the on shell momentum of the $i$th line in this loop, of momentum $k_i$.   Neglecting external
momenta and masses for large loop momenta, we have
\bea
[k_i]^- = - k_i^\tm\, \frac{k_i^\tp}{k^+} =  - \left(\eta_{i b} l_b^\tm+ \sum_{a\ne b} \eta_{i a} l_a^\tm\right)\, \frac{ \left( \eta_{i b} l^\tp+ \sum_{a\ne b} \eta_{i a} l_a^\tp\right)}{k_i^+}\, ,
\eea
where as in Eq.\ (\ref{eq:ki-sum}), $\eta_{bi}=\pm 1$ around the loop, depending on whether loop $l_b$ flows with or against the 
defining direction of line momentum $k_i$, and
where the sum over $a$ includes all loop momenta with the exception of $l_b$.
To be definite, suppose $l_b^\tp$ is large and positive.
The condition that each component $k_i^\tp$ flows in the same direction as loop momentum $l_b$ can then be written as
\bea
 \eta_{ib} l_b^\tp+ \sum_{a\ne b} \eta_{ i a} l_a^\tp   =\eta_{i b} \left |   l_b^\tp+ \frac{1}{\eta_{i b}} \sum_{a\ne b} \eta_{ i a} l_a^\tp\right |
 \, .
\eea
We then have 
\bea
[k_i]^- = - \left( \eta^2_{ i b} l_b^\tm+ \eta_{i b}\, \sum_{a\ne b} \eta_{ i a} l_a^\tm\right)\, 
\frac{\left |   l_b^\tp+ \frac{1}{\eta_{i b}} \sum_{a\ne b} \eta_{ i a} l_a^\tp\right |}{k_i^+}\, ,
\eea
and the coefficient of $l_b^\tm$ is always positive for every term in which it appears in the LCOPT denominators of Eq.\ (\ref{eq:lcopt}),
since $k_i^+$ is also always positive. 
All $l_b^\tm$ poles are thus in the same half plane (lower for $l_b^\tp$ positive), and the integrals vanish so long as the loop appears in at least two states.   This, however, is ensured by our assumption of an ultraviolet-finite scalar diagram.

\subsection{Unitarity(ies)}

\begin{figure}
\centerline{\figscale{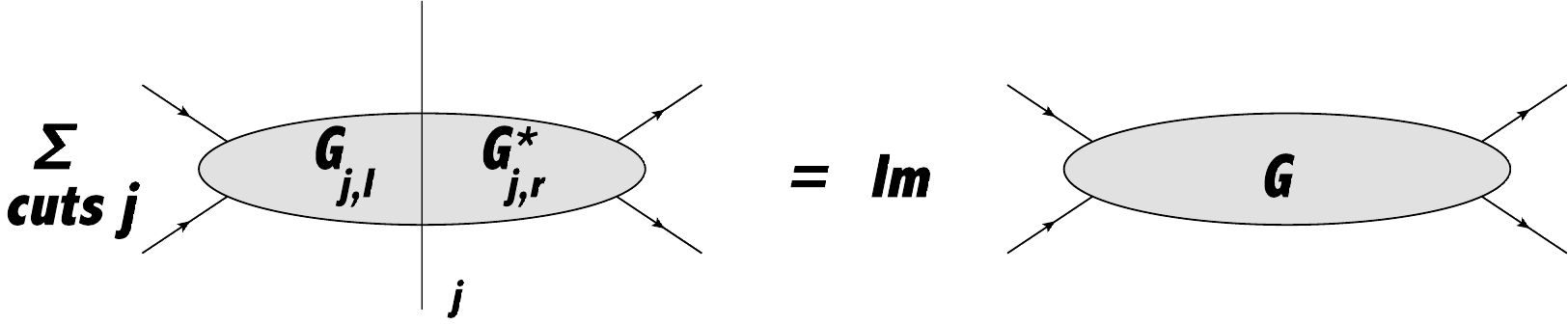}{14cm}}
\caption{A representation of perturbative unitarity, Eq.\ (\ref{eq:unitarity_cut}) for an
arbitrary diagram $G$.   As shown in the text, after an integral over loop
$l^-$ integrals, this relation holds for each light cone
ordering of diagram $G$ at fixed values of all loop $l^+$, and $l^1,l^2$ or
$l^\tp$ and $l^\tm$ \label{fig:unitarity}. A similar result holds when all $l^\tm$ integrals are carried out
at fixed $l^\tp$, $l^+$ and $l^-$.}
\end{figure}

The light cone ordered expression (\ref{eq:lcopt}) for an arbitrary diagram implies
the order-by-order unitarity of perturbation theory, a relation that has been used extensively in showing the cancellation of
infrared divergences in inclusive cross sections \cite{Sterman:1978bi,Lee:1964is,Collins:1989gx}.   Here we note only the fundamental identity
at the basis of this connection.   We consider an arbitrary diagram ${\cal G}^{(T)}$,
with a specific light cone order $T$, and sum over the terms found by setting each state, $s_i$ of $T$ on shell in turn,
replacing its light cone denominator by a delta function.    
Each such substitution  we refer to as a ``cut" of the diagram.  All states before (to the left of) the cut retain
a $+i\ep$ prescription, and those after the cut (to the right) are given a $-i\ep$ prescription.  See the left hand side of Fig.\ \ref{fig:unitarity}.

Each cut in the figure splits the ordered diagram into two ordered sub-amplitudes, ${\cal G}_{j,l}^{(T)}$ and ${\cal G}_{j,r}^{(T)}$,
at fixed loop momenta
to the ``left" and ``right" of the cut, respectively.    The fundamental identity, which holds at fixed values
of the all loop momenta $l_a^+$, $l_a^\tp$ and $l_a^\tm$, is
\begin{eqnarray}
 {\cal G}_{j,r}^{(T)}{}^*{\cal G}_{j,l}^{(T)} &=& \sum_{j=1}^{V_{\cal G}-1}\ \left(\prod _{i'=j+1}^{V_{\cal G}-1}
\frac{1}{P_{i'}^- - s_{i'}-i\epsilon}\right) 2\pi\delta\left(P_j^- - s_{j}\right) \left(\prod _{i=1}^{j-1} \frac{1}{P_i^- - s_{i}+i\epsilon}\right)
\nn\\
&=& -i\, \left[
\left(\prod _{i'=1}^{V_{\cal G}-1}
\frac{1}{P_{i'}^- - s_{i'}-i\epsilon}\right) \ -\ 
\left(\prod _{i=1}^{V_{\cal G}-1}  \frac{1}{P_i^- - s_{i}+i\epsilon}\right) \right]
\nn\\
&= & -i\, \left[ {\cal G}^* - {\cal G} \right]\, ,
\label{eq:unitarity_cut}
\end{eqnarray}
where $\cal G$ is the uncut diagram at fixed remaining components of loop momenta and $V_{\cal G}$ the number of vertices in $\cal G$.
The on shell value of minus momentum for state $i$ is $s_i$.
The proof of this relation follows easily from repeated use of the distribution identity,
$2\pi i \delta(x) = 1/(x-i\ep) - 1/(x+i\ep)$.
In this form the integrand of the sum of cut diagrams is related to the imaginary part of the integrand 
for the uncut diagram, a generalized form of the optical theorem, as
illustrated by Fig.\ \ref{fig:unitarity}.

At the level of the fundamental identity, Eq.\ (\ref{eq:unitarity_cut}), then, unitarity is
a property of perturbation theory in $(2,2)$ signature as much as in Minkowski space.   In fact, we can 
derive light cone ordered perturbation theory just as well by performing 
the $l^\tm$ integrals as the $l^-$ integrals, deriving an identity 
of exactly the same form as Eq.\ (\ref{eq:unitarity_cut}) for an arbitrary 
diagram, but now at fixed loop momenta $l^+$, $l^-$ and $l^\tp$.  
In a sense, then, there is an extra unitarity relation for $(2,2)$ compared to $(1,3)$.
We do not have a practical application of this result to propose at this time.

\section{$2L$-dimensional representation}
\label{sec:covariant-rep}

The double set of  light cone coordinates
of Eq.\ (\ref{eq:kdefs}) can be used to derive a new representation for diagrammatic integrals,
based on the linearity of all propagators in the minus and minus tilde variables.
We start with the general scalar integral,
Eq.\ (\ref{eq:I_LN_22}), in $(2,2)$ signature
 for an arbitrary diagram with $L$ loops and $N$ lines, assuming that $L_g > 2N_g$ for any subgraph, $g$,
so that all subintegrals are convergent,
\bea
I _{N,L}(p_j) \equiv -\ \int \prod_{i=1}^L \frac{dl_i^+ \, dl_i^\tp}{2(2\pi)^2}\, \frac{dl_i^\tm \, dl_i^-}{2(2\pi)^2}\, \prod_{\alpha =1}^N \frac{1}{D_\alpha}\, .
\label{eq:I-original}
\eea
In the defining normalizations of Eq.\ (\ref{eq:kdefs}), the denominators are given by
\bea
D_\alpha &=&  \left( l_\alpha-p_\alpha \right)^2-m_\alpha^2 + i\ep 
\nn\\
&\equiv& \left (l_\alpha^+ - p_\alpha^+ \right ) \left (l_\alpha^- - p_\alpha^- \right )+
\left (l_{\alpha}^\tp -p_{\alpha}^\tp \right ) \left (l_{\alpha}^\tm - p_{\alpha}^\tm  \right ) - m_\alpha^2 +i\ep\, .
\label{eq:D-explicit}
\eea
Here $l_\alpha$ and $p_\alpha$ are the combinations of loop momenta $l_i$ and external momenta $p_j$, respectively, flowing along 
internal line line $\alpha$, with momentum $k_\alpha$.    In the notation of Eq.\ (\ref{eq:ki-sum}),
\begin{align}
l_\alpha=\eta_{\alpha i}l_i\, , \quad p_\alpha = \xi_{\alpha j}p_j\, ,
\end{align}
with $\eta_{\alpha i},\ \xi_{\alpha j}=\pm 1,0$.   
Making the minus and minus-tilde loop momentum dependence explicit, we write the denominators as
\begin{align}
D_\alpha=A_{\alpha i}^+ l_i^-\ +\ A_{\alpha i}^\tp l_{i}^\tm+B_\alpha\, ,
\end{align}
in terms of coefficients $A$ and  $B$, defined by
\bea
A_{\alpha i}^+ &=& (l_\alpha^+ - p_\alpha^+)\eta_{\alpha i}\, ,  \nonumber \\
A_{\alpha i}^\tp  &=& (l_{\alpha}^\tp - p_{\alpha}^\tp )\eta_{\alpha i}\, ,  \nonumber \\
B_\alpha &=&  (p_\alpha^+ - l_\alpha^+)p_\alpha^-\ +\   (p_{\alpha}^\tp - l_{\alpha}^\tp  )p^\tm_{\alpha} - m_\alpha^2
\nn\\
&=& p_\alpha^2 - m_\alpha^2 - 2\hat p_\alpha \cdot l\, ,
\label{eq:AB-coeff-def}
\eea
where in the second relation for $B_\alpha$, we define a vector with only minus and minus tilde components,
\bea
\hat p_\alpha^\mu \equiv \left (0^+, p_\alpha^-,0^\tp, p_\alpha^\tm\right)\, .
\label{eq:hat-pdef}
\eea
The linearity of all denominators, (\ref{eq:D-explicit}) in both sets of integration variables
 $\{l_i^-\}$ and $\{l_i^\tm\}$ will allow us to derive an explicit form
for each integral $I_{N,L}$ as a sum over choices of $2L$ on shell (`cut')  lines.   

Our  integrals can be put into a more compact form by introducing a single index to cover the sum
over components {\it and} loops,
\bea
I_{N,L}(p_j)= -\
 \left( \frac{1}{4(2\pi)^4}\right)^L\
\int \prod_{k=1}^{2L} dy_k \int \frac{\prod_{j=1}^{2L} dx_j}{\prod_{\alpha=1}^N (\sum_{j=1}^{2L} A_{\alpha j}x_j+B_\alpha+i\epsilon)}\, ,
\label{eq:double-lc-integral}
\eea
where  $\{x_j\}\equiv \{l_i^-,l_i^\tm\}$, runs over
the minus and minus tilde components of all loops and $\alpha$ over the set of lines.  
To make our result as explicit as possible, we are free to define
\bea
x_{2i-1} &=& l_i^-\, ,
\nn\\
x_{2i} &=& l_i^\tm\, ,
\label{eq:xj-def}
\eea
where $i$ runs from $1$ to $L$.
Correspondingly, we may define the remaining $2L$ integration variables as
\bea
y_{2i-1} &=& l_i^\tp\, ,
\nn\\
y_{2i} &=& l_i^+\, ,
\label{eq:yi-def}
\eea
for the set $y_k$.   The relabeled coefficients
$A_{\alpha i}$ are then linear functions of parameters $y$ and can
be thought of as defining a matrix.   To be explicit, in terms of the coefficients
of Eq.\ (\ref{eq:AB-coeff-def}), we define
\bea
A_{\alpha,2i-1} &\equiv& A_{\alpha i}^+ \, ,
\nn\\
A_{\alpha,2i} &\equiv& A_{\alpha i}^\tp\, . 
\eea
We may choose to do the
integrals in the order $y_1 \cdots y_{2L}$, and 
as we will see, individual terms in our results
depend in a structured manner on the order of integration.   
The final result, however, cannot depend on the order.

 The essential observation regarding the integral in Eq.\ (\ref{eq:double-lc-integral}) is that 
the singularity structure of the integrand for each $x_j$
is simple poles at every step in the integration procedure, and  that closing on these poles does not affect the limits of the remaining $x_j$, only the $y_j$.   
We will choose to perform these integrals by closing contours in each lower half complex $x_j$-plane.
The choice of each pole sets one line on shell, and at the end of $2L$ integrations we have
a sum of terms in which $2L$ lines are ``cut" in this fashion.    Let an arbitrary sequence of $k$ lines found
in this way be labelled $\ca_{k}$, where $k=1$ labels the first line set 
on shell, and $\ca_{2L}$ the full set for the sequence.   Each set $\ca_{k}$ must be such that: (i) its lines carry $k$
linearly independent loop momenta, and (ii) after any $m$ integrals $x_1 \dots x_m$, $m\le k-1$,
there must be a lower half-plane pole in the next integration variable, $x_{m+1}$.
Let us denote by $A^{({\ca_k})}$ the $k\times k$ matrix whose elements are $A_{\alpha j}$,
such that $j=1 \dots k$ and $\alpha \in \ca_k$.

The result we are after clearly depends on
the values of the $x_j$  when $k$, $k=1 \dots 2L$, lines are set on shell, that is on solutions to a system of
$2L$ linear equations in $2L$ variables.   For any choice of $k$ lines, where $k$ need not be
an even number, these equations are 
\bea
A^{(\ca_{k})}_\alpha \cdot x +B^{(\ca_{k})}_\alpha + i\ep \equiv
\sum_{j=1}^{k} A^{(\ca_{k})}_{\alpha j} x_j+B^{(\ca_{k})}_\alpha + i\ep=0\, , \quad \alpha \in \ca_{k}\, ,
\label{eq:linear-equations}
\eea
where, again, the superscripts identify $A^{(\ca_{k})}$ as a $k\times k$ matrix and $B^{(\ca_{k})}$
as a $k$-component vector.   
The matrix, of course, must be non-singular, which is to say that we will
find $k$ independent poles only if the momenta of these lines are linearly independent.
The solution to Eq.\ (\ref{eq:linear-equations})
can be represented in terms of its real and imaginary parts $x_j=X^{({\cal A}_{k})}_j+i\ep Y^{({\cal A}_{k})}_j$, $j=1\dots k$ as
\footnote{Here we assume that all masses are real.   The generalization to masses with positive imaginary parts is immediate.}
\bea
X^{({\cal A}_{k})}_j &=&  -\ \sum_{\alpha'}\ \left(A^{({\cal A}_{k})}\right)^{-1}_{j\alpha'} B^{({\cal A}_{k})}_{\alpha'}\, ,
\nn\\
Y^{({\cal A}_{k})}_j &=&  -\  \sum_{\alpha'}\ \left(A^{({\cal A}_{k})}\right)^{-1}_{j\alpha'}\, ,
\label{eq:solns}
\eea
in terms of the inverse of matrix $A^{({\cal A}_{k})}$.   Note the sum over unrepeated index $\alpha'$ in the expression
for the imaginary part.  
The solutions in (\ref{eq:solns}) determine the values of the remaining denominators when all $k \rightarrow 2L$ 
denominators are replaced by delta functions.   This result alone does not determine the integral, however, 
because of theta functions that result from closing each contour in the lower half-plane
 in turn.   The arguments of these step functions depend, in general, on the order in which
 the integrals are carried out.

We will now show that in the notation of Eq.\ (\ref{eq:solns}), the result of doing the $2L$ $x_j$ integrals in  (\ref{eq:double-lc-integral}) is given by
\bea
I_{N,L} &=&  -\ \left( \frac{-1}{4(2\pi)^2}\right)^L\
\sum_{{\cal A}_{2L}} \int \prod_{k=1}^{2L} dy_k\  
\theta\left( \frac{\det A^{(\ca_{k-1})}\; F^{(\ca_k)}_{\alpha_k}(y_1\dots y_k)}{\det A^{(\ca_k)}(y_1\dots y_k)}\right)
\nn\\
&\ & \hspace{10mm} \times\
 \frac{1}{\det (A^{(\ca_{2L}) })}\frac{1}{\prod_{\beta \notin \ca_{2L}}(A_\beta \cdot X^{(\ca_{2L})}+B_\beta+i\epsilon(1+A_{\beta } \cdot Y^{({\ca }_{2L})}))}\, .
 \label{eq:L-N-2L}
\eea
The product of theta functions depends, as suggested above, on the order of integration.
For the $k$th integration, we find
\begin{align}
F^{(\ca_k)}_{\alpha_k} =  1+ \sum_{j=1}^{k-1}A^{(\ca_{k})}_{\alpha_k j}\, Y_j^{({\cal A}_{k-1})} \,  ,
\label{eq:Fk-def}
\end{align}
 where $\alpha_k$ is the index of the $k$th line put on shell, as above $A^{(\ca_k)}$ is the $k\times k$ matrix associated with the
 first $k$ lines, and
where $Y_j^{({\cal A}_{k-1})}$ is  the solution for the imaginary part of $x_j$ given in (\ref{eq:solns})
when the first $k-1$ lines are put on shell.    
It should be noted that in the sum over sequences ${\cal A}_{2L}$ 
there are many terms that differ only in sign and integration region.  The sign comes from the determinant of $A^{(\ca_{2L})}$. 
Note the response of the imaginary parts to the selection of poles, as analyzed in the context of
``loop-tree" dualities for Minkowski integrals \cite{Catani:2008xa,Bierenbaum:2010xg,CaronHuot:2010zt}.

For an inductive proof of Eq.\ (\ref{eq:L-N-2L}), we start by noting that that the role of the $y_j$ is entirely passive.   We 
need therefore only consider the proof of 
\bea
J_{N,l}\left (A_{\alpha i},B_{\alpha}\right) &\equiv& \int \frac{\prod_{j=1}^{l} dx_j}{\prod_{\alpha=1}^N (\sum_{j=1}^{l} A_{\alpha j}x_j+B_\alpha+i\epsilon)}
\nn\\
&=& -\ (-2\pi i)^{l}\sum_{{\cal A}_{l}}  \prod_{k=1}^{l} \  
\theta\left( \frac{\det A^{(\ca_{k-1})}\; F^{(\ca_k)}_{\alpha_k}(y_1\dots y_l)}{\det A^{({\cal A}_k)}(y_1\dots y_l)}\right)
\nn\\
&\ & \hspace{10mm} \times\
 \frac{1}{\det (A^{(\ca_{l}) })}\frac{1}{\prod_{\beta \notin \ca_{l}}(A_\beta \cdot X^{(\ca_{l})}+B_\beta+i\epsilon(1+A_{\beta} \cdot Y^{({\ca}_l)} ))}\, ,
 \nn
 \\
 \label{eq:k-hypothesis}
\eea
for arbitrary $l$. The case of $l=1$, $J_{N,1}$ is easily verified, and for any $l$, we can 
use the relation
\bea
J_{N,l}\left (A_{\alpha i},B_{\alpha}\right) =  \int dx_{l}\  J_{N,l-1} \left (A_{\alpha i},B_{\alpha}+ A_{\alpha l} x_{l} \right)\, ,
\eea
in which the $x_l$ integral of $J_{N,l}$ is absorbed into the $B$'s for $J_{N,l-1}$.   Now assuming the result (\ref{eq:k-hypothesis})
for $l-1$, and using (\ref{eq:solns}), we have 
\bea
J_{N,l}\left (A_{\alpha i},B_{\alpha}\right) 
&=& -\ (-2\pi i)^{l-1}\ \int dx_{l} \ \sum_{{\cal A}_{l}}  \prod_{k=1}^{l-1} \  
\theta\left( \frac{det A^{(\ca_{k-1})}\; F_{\alpha_k}(y_1\dots y_k)}{\det A^{({\cal A}_k)}(y_1\dots y_k)}\right)
 \frac{1}{\det (A^{(\ca_{l}) })}
\nn\\
&\ & \hspace{-5mm} \times\
\prod_{ \beta \notin \ca_{l-1}} 
\Bigg [ \left( A_{\beta l}\ -\ A_{\beta j}  \left(  A^{({\cal A}_{l-1})}\right)^{-1}_{j\alpha'} A_{\alpha' l} \right) x_{l}
\nn\\
&\ & \hspace{15mm} 
+ \ B_\beta
-\ A_{\beta j}  \left(  A^{({\cal A}_{l-1})}\right)^{-1}_{j\alpha'}B^{({\cal A}_{l-1})}_{\alpha'} 
+i\epsilon \left (1\ -\ A_{\beta j}\sum_{\alpha'} \left(A^{({\cal A}_{l-1})}\right)^{-1}_{j\alpha'} \right ) \Bigg ]^{-1}\, .
\nn\\
 \label{eq:k-l+1}
\eea
To this expression, we apply an elementary identity, applicable to any nonsingular, $(n+1) \times (n+1)$ matrix, $M^{(n+1)}$ defined by $M_{i,j}$,
$i,j = 1 \dots n+1$ in terms of its submatrix $M^{(n)}_{a,b} \equiv M_{a,b}$, $a,b = 1 \dots n$, 
\bea
\frac{ \det M^{(n+1)} }{\det M^{(n)} }
= M_{n+1, n+1}\ -\  \sum_{i=1}^{n} \sum_{j=1}^{n} M_{n+1, i} \left( M^{(n)} \right)^{-1}_{i,j} M_{j , n+1}\, .
\label{eq:matrix-identity}
\eea
This is readily proved using the relation of the inverse of a matrix to minors of its determinant.
Applying Eq.\ (\ref{eq:matrix-identity}) to the coefficient of $x_{l}$
in (\ref{eq:k-l+1}), the form of Eq.\ (\ref{eq:k-hypothesis}) for $J_{N,l}$ is 
then simply the sum of residues found by closing the $x_{l}$ integral in the lower half plane.
By identifying $l$ with $2L$, Eq.\ (\ref{eq:L-N-2L}) follows directly.

In fact, the identity (\ref{eq:matrix-identity}) can be applied again, to the imaginary and real parts of (\ref{eq:L-N-2L}),
to provide an alternative expression for the integrand in eq.\ (\ref{eq:L-N-2L})
entirely in turns of the matrices $A_{\alpha i}$ and 
vectors $B_\alpha$.    For each sequence $\ca_{k}$, we find in the remaining denominators, $\beta$,
\bea
A_\beta \cdot X^{(\ca_{k})}+B_\beta
=
\frac{1}{\det A^{(\ca_k)}} \left| 
\begin{matrix}
A^{(\ca_k)}_{\alpha_1 1} & \cdots  A^{(\ca_k)}_{\alpha_1 n} & B_{\alpha_1} \\
\vdots        &  \vdots &        \vdots  \\
A^{(\ca_k)}_{\alpha_k 1} & \cdots  A^{(\ca_k)}_{\alpha_k k} & B_{\alpha_n} \\
A_{\beta 1} & \cdots  A_{\beta k} & B_{\beta} 
\end{matrix} 
\right|
\ \equiv\ \frac{G_\beta^{( \ca_{k+1})}}{\det A^{(\ca_k)}}\, .
\label{eq:Gdef}
\eea
We have a similar form for the arguments of the theta functions in Eq.\ (\ref{eq:L-N-2L}),
\bea
F^{(\ca_{k+1})}_{\beta} =
\frac{1}{\det A^{(\ca_{k})} }\left| 
\begin{matrix}
A^{(\ca_{k})}_{\alpha_1 1} & \cdots  A^{(\ca_{k})}_{\alpha_1 k-1} & 1 \\
\vdots        &  \vdots &        \vdots  \\
A^{(\ca_{k})}_{\alpha_{k} 1}  & \cdots  A^{(\ca_{k})}_{\alpha_{k} k} & 1\ \\
A_{\beta 1} & \cdots  A_{\beta k} & 1 
\end{matrix} 
\right|\ \equiv\ \frac{H_\beta^{ ( \ca_{k+1})}}{\det A^{(\ca_{k})}}\, .
\label{eq:Hdef}
\eea
We can thus reinterpret the result of the $x_i$ integrals, Eq.\ (\ref{eq:L-N-2L}) as
\bea
I_{N,L} &=&  -\ \left( \frac{-1}{4(2\pi)^2}\right)^L\
\sum_{{\cal A}_{2L}} \int \prod_{k=1}^{2L} dy_k\  
\theta\left( \frac{H_{\alpha_k}^{(\ca_{k})}}{\det A^{({\cal A}_k)}}\right)
 \left( \det A^{(\ca_{2L}) } \right )^{N-2L-1}\
 \nn\\
 &\ & \hspace{40mm} \times\
\prod_{\beta \notin \ca_{2L} } \frac{1}{ G_\beta^{(\ca_{2L+1})} + i\epsilon H_\beta^{(\ca_{2L+1})}  } 
 \, ,
 \label{eq:alt-L-N-2L}
\eea
where, as the notation indicates, the determinants $G$ and $H$ are of $(2L+1)\times (2L+1)$ matrices,
determined in each case by the coefficients   of on shell lines, and of each remaining, uncut line $\beta$.
In this expression
 the entire integrand is specified by determinants of elements $A_{\alpha i}$ and $B_\alpha$.   
 These coefficients, in turn, given
in (\ref{eq:AB-coeff-def}), are linear functions of the plus 
and plus tilde loop momentum components in addition to external momenta and masses.  
Note that for $k=1$,  the theta function corresponds to the condition that the pole in the first integral, 
over loop momentum $l_1^-$, be in the lower half-plane,
so that, because the set $\ca_1$ consists of one line only, say $i$, we have
\bea
H^{(A_1)}_{\alpha_1}  &\equiv& 1\, ,
\nn\\
\det A^{(A_1)} &=& \left(l_i^+ - p_i^+ \right)\eta_{i1}  \, ,
\label{eq:HA_1}
\eea
with no sum on $i$ in the second expression.
The integrand in Eq.\ (\ref{eq:alt-L-N-2L})
is a rational function of the remaining $2L$ components, $y_j$.   Individual denominators
labelled by index $\beta$ may involve powers of up to order
$2L+1$ in these variables, although by examining the one-loop case below, we will see that the power can be lower.

Eq.\ (\ref{eq:alt-L-N-2L}) is our final result for ultraviolet finite scalar integrals in $(2,2)$.  
For any such diagram, $2L<N-1$, so that the number of integrations remaining is fewer than
the number of Feynman parameter integrals for the corresponding diagram, at the price of having a sum of terms.
In these expressions, the finiteness of the remaining integration regions, shown in
Sec.\ \ref{sec:finite} above, is not manifest.   It results from cancellations between different
terms at each stage in the integration.   We will give an example in the next
section, where we study the one-loop case.

\section{One Loop Diagrams}
\label{sec:one-loop}

We now turn to the application of our basic result,  (\ref{eq:alt-L-N-2L}) to one loop diagrams.    We begin
with a one loop diagram of any order, with completely arbitrary real masses and external momenta.   
We will not attempt to perform the remaining two integrals, but will be able to identify certain
interesting general features.  Following this, we confirm the presence of double-logarithmic behavior
in a sample $(2,2)$ box diagram.

\subsection{The general one loop diagram in (2,2) notation}

For the case $L=1$ in Eq.\ (\ref{eq:alt-L-N-2L}), the sum over sets of cut lines, $\ca_1$ and $\ca_2$ is simply 
a sum of ordered choices of lines, say $\alpha_1=i$ and $\alpha_2=j$, which we will denote
by $\ca_1=\ca_i$ and $\ca_2=\ca_{(ij)}$.   With the labeling of momenta 
specified in Eq.\ (\ref{eq:AB-coeff-def}),  the first
index, $\alpha_1=i$ denotes the line set on shell by the integral over loop component $x_1=l^-$, while $\alpha_2=j$ 
identifies the line set on shell by the integral over $x_2=l^\tm$, in the notation of Eq.\ (\ref{eq:xj-def}).
  In these terms, we find, using (\ref{eq:HA_1}), for $L=1$,
\bea
I_{N,1} &=& \frac{1}{4(2\pi )^{2}} \, \sum_{i,j} \int  dl^+\  \theta\left( \frac{1}{l^+-p_i^+}\right)
\int dl^\tp \theta\left( \frac{H_{\alpha_j}^{(\ca_{(ij)})}}{\det A^{({\cal A}_{ij})}}\right)
\nn\\
&\ & \hspace{20mm} \times 
 \left( \det A^{(\ca_{(ij)}) } \right )^{N-3}\
\prod_{\beta \ne i,j} \frac{1}{ G_\beta^{(\ca_{(ij\beta)})} + i\epsilon H_\beta^{(\ca_{(ij\beta)})}  } \, ,
 \label{eq:alt-2-N-2}
\eea
where $\ca_{(ij\beta)}$ in the superscripts of determinants $G$ and $H$ corresponds to $\ca_{2L+1}$ in (\ref{eq:alt-L-N-2L}).
To illustrate the method, we evaluate the remaining determinants in the expression.
These are from the $2\times 2$ matrices, $A^{(\ca_{(ij)})}$,
\bea
\det A^{(\ca_{(ij)})} &=&
\left | 
\begin{matrix}
l^+-p_i^+ &  l^\tp - p_i^\tp \\
l^+-p_j^+ &  l^\tp - p_j^\tp
\end{matrix}
\right |
\ =(\l^+ - p_i^+)(p_i^\tp - p_j^\tp ) - (l^\tp - p_i^\tp) ( p_i^+ - p_j^+)\, ,
\eea
\label{eq:A-oneloop}
and  $H_{\alpha_j}^{(\ca_{(ij)})}$,
\bea
H_{\alpha_j}^{(\ca_{(ij)})}
&=&
\left | 
\begin{matrix}
l^+-p_i^+ &  1 \\
l^+-p_j^+ &  1
\end{matrix}
\right |
\ =\ p_j^+ - p_i^+\, ,
\label{eq:H-thetaoneloop}
\eea
and the two $3\times 3$ matrices, $G_\beta^{(\ca_{(ij)})}$,
\bea
\det G_\beta^{(\ca_{(ij)})} &=& 
\left | 
\begin{matrix}
l^+-p_i^+ &  l^\tp - p_i^\tp & B_{i} \\
l^+-p_j^+ &  l^\tp - p_j^\tp & B_{j} \\
l^+-p_\beta^+ &  l^\tp - p_\beta^\tp & B_{\beta} 
\end{matrix} 
\right |
\nn\\
&=& B_i\, \det A^{(\ca_{(j\beta)})}\ - B_j\, \det A^{(\ca_{(i\beta)})}\ +\ B_\beta\, \det A^{(\ca_{(ij)})}\, ,
\label{eq:G-oneloop}
\eea
and $H_\beta^{(\ca_{(ij)})}$,
\bea
\det H_\beta^{(\ca_{(ij)})} &=&
\left | 
\begin{matrix}
l^+-p_i^+ &  l^\tp - p_i^\tp & 1 \\
l^+-p_j^+ &  l^\tp - p_j^\tp & 1 \\
l^+-p_\beta^+ &  l^\tp - p_\beta^\tp & 1 
\end{matrix} 
\right |
\ =\
(p_\beta^+ - p_i^+)(p_\beta^\tp - p_j^\tp) - (p_\beta^+ - p_j^+)(p_\beta^\tp - p_i^\tp) \, .
\nn\\
\label{eq:H-oneloop}
\eea
Recalling that the $B_i$ are linear in loop momenta, we see that the denominators
$\beta$ in Eq.\ (\ref{eq:alt-2-N-2}) are of power two jointly in $l^+$ and $l^\tp$,
rather than three.   

In order to write our result in a more compact form, we introduce an antisymmetric product
\bea
\{ v,w \} &\equiv& v^+ w^\tp -  w^+ v^\tp\, .
\label{eq:bracket-def}
\eea
In this notation, the general one-loop scalar integral becomes
\bea
I_{N,1} &=& \frac{1}{4(2\pi )^{2}} \sum_{i,j} \int  dl^+\  \theta\left( l^+-p_i^+ \right)
\int d l^\tp \theta\left( \frac{ \{ l, p_i-p_j \}  + \{p_i,p_j\} }{p_j^+-p_i ^+} \right)\  \left( \{ l, p_i-p_j \} + \{p_i,p_j\}\right )^{N-3} 
\nn\\
&\ & \hspace{15mm} \times 
\prod_{\beta \ne i,j} \frac{1}{ \frac{1}{2} \, \sum_{\{a,b,c\} =\{i,j,\beta\}}  \ep_{abc} B_a \left( \{l, p_b - p_c\} +\{p_b,p_c\} \right)
+i\ep\{p_\beta-p_i,p_\beta-p_j\}  }
 \nn\\
 &\equiv& \frac{1}{4(2\pi )^{2}} \sum_{i,j} \int  dl^+\  \theta\left( l^+-p_i^+ \right)
\int dl^\tp \theta\left(  l^\tp - l^+r_{p_i-p_j}  +\frac{ \{p_i,p_j\} }{p_j^+-p_i ^+} \right)\ \omega_{ji}(l^+,l^\tp)\, ,
 \label{eq:alt-2-N-3}
\eea
where in the second equality we have evaluated the theta function for the $l^\tp$ integral, 
using the notation of Eq.\ (\ref{eq:mudefs}), and have
defined $\omega_{ji}$ as the integrand that results from taking the $i$th pole for the $l^-$
integral, and the $j$th pole for $l^\tm$.   The $\epsilon$ symbol reflects the antisymmetry
of determinant $G_\beta^{(\ca_{(ij)})}$, Eq.\ (\ref{eq:G-oneloop}), in indices $i$, $j$ and $\beta$,
corresponding to line momenta $k_i$, $k_j$ and $k_\beta$.
We now note that for fixed $i$, the subsequent $l^\tp$ integral vanishes if all its poles are
in the lower half plane, which leads to the identity,
\bea
\sum_{j\ne i} \omega_{ji}(l^+,l^\tp) = 0\, .
\label{eq:sum-zero}
\eea
This enables us to rewrite $I_{N,1}$, (\ref{eq:alt-2-N-2}) as 
\bea
I_{N,1} &=& \frac{1}{4(2\pi )^{2}} \sum_i \int_{p_i^+}^\infty dl^+ \sum_{j\ne i}\ 
\left( \int_{l^+r_{p_i-p_j}  - \frac{ \{p_i,p_j\} }{p_i^+-p_j ^+}}^{\infty}
d l^\tp \ \omega_{ji}(l^+,l^\tp) \ - \  \int^{\infty}_{\sigma(l^+)} dl^\tp \omega_{ji}(l^+,l^\tp) \right)
\nn\\
&=& \frac{1}{4(2\pi )^{2}} \sum_i \sum_{j\ne i}\ \int_{p_i^+}^\infty dl^+   \int_{l^+r_{p_i-p_j}  - \frac{ \{p_i,p_j\} }{p_i^+-p_j ^+}}^{\sigma(l^+)}
d l^\tp \ \omega_{ji}(l^+,l^\tp)\, ,
\nn\\
\label{eq:I-2-N-3a}
\eea
where $\sigma(l^+)$ is a completely arbitrary function of $l^+$ (possibly a constant), which must be chosen the
same for every pair $i,j$.

\begin{figure}
\centerline{\figscale{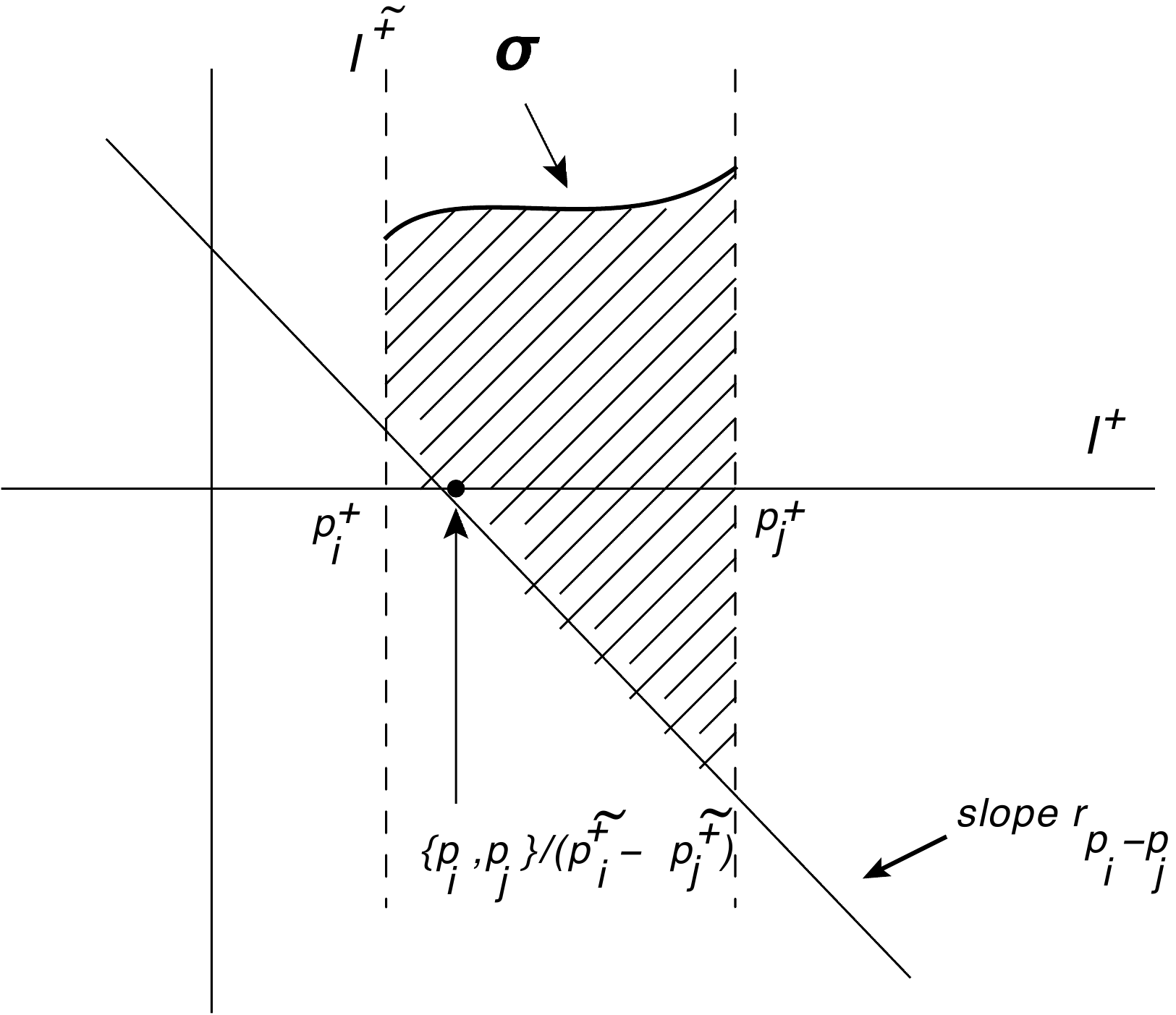}{8cm}}
\caption{Region of integration (shaded) corresponding to Eq.\ (\ref{eq:alt-2-N-4}) in the $l^+,l^\tp$ plane.   For the case shown,
the parameter $r_{p_i-p_j}$ is negative, corresponding to a negative slope in the lower limit of the $l^\tp$ 
integral.   Positive slopes and negative intercepts are also possible.   As explained in the text, the boundary $\sigma$ is arbitrary.
\label{fig:theta-fns}}
\end{figure}

We can simplify this expression further by using that in Eq.\ (\ref{eq:I-2-N-3a}), the integrand $\omega_{ji}$ is fully 
antisymmetric under the exchange of  $p_i$ and $p_j$, that is,
\bea
\omega_{ji}(l^+,l^\tp) = - \omega_{ij}(l^+,l^\tp)\, .
\label{eq:antisym}
\eea
Equation (\ref{eq:I-2-N-3a}) can thus
be rewritten as a sum over $(1/2)N(N-1)$ ordered pairs of terms, with fixed limits on the $l^+$ integrals,
and linear one-sided limits for the $l^\tp$ integrals, 
\bea
I_{N,1} &=& \frac{1}{4(2\pi )^{2}} \sum_{i} \sum_{j\ne i} \theta(p_j^+-p_i^+) \int_{p_i^+}^{p_j^+}  dl^+\  
\int^{l^+r_{p_i-p_j}  - \frac{ \{p_i,p_j\} }{p_i^+-p_j ^+}}_{\sigma(l^+)}
d l^\tp\   
\nn\\
&\ & \hspace{5mm} \times 
\frac{\left( \{ l, p_i-p_j \} + \{p_i,p_j\}\right )^{N-3}}{ \prod_{\beta \ne i,j} \left[ \frac{1}{2} \, \sum_{\{a,b,c\} =\{i,j,\beta\}} \ep_{abc} 
B_a \left( \{l, p_b - p_c\} +\{p_b,p_\beta\} \right)
+i\ep\{p_\beta-p_i,p_\beta-p_j\} \right]}\, ,
\nn\\
 \label{eq:alt-2-N-4}
\eea
where we observe again that because of the identity (\ref{eq:sum-zero}), the result is independent of our choice of $\sigma(l^+)$.
The integration region is illustrated in Fig.\ \ref{fig:theta-fns}.

In principle, Eq. (\ref{eq:alt-2-N-4}) could be the starting point of an explicit calculation, but in any
case an arbitrary one-loop diagram can be reduced to box diagrams \cite{Brown:1952eu}-\cite{Bern:1993kr},
which are known for any choices of masses \cite{'t Hooft:1978xw}-\cite{Denner:2010tr}.    Our emphasis here 
is rather on the extension of the formalism to the new signature.

\subsection{Double logs in a (2,2) box}

We have already argued that four-point amplitudes are insensitive to the choice
of Minkowski or $(2,2)$ signature.   To illustrate this point, let us show how
double-logarithmic integrals arise in
 the one-loop box with a suitable choice of massless internal and external lines, 
 directly from the $(2,2)$ result, Eq.\ (\ref{eq:alt-2-N-4}) with $N=4$.
  
\begin{figure}
\centerline{ \figscale{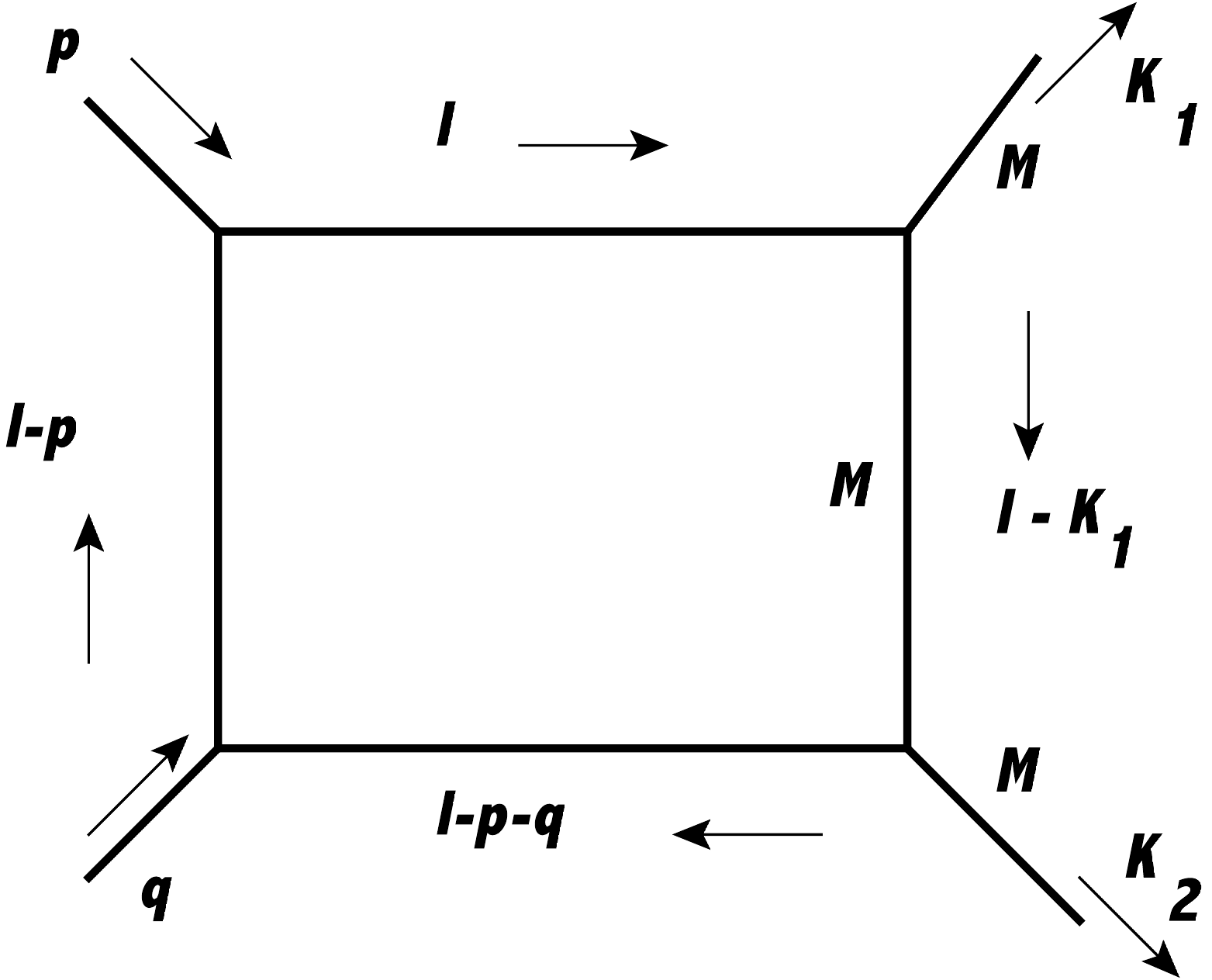}{8cm}}
\caption{Box diagram}
\label{fig:box}
\end{figure}

We consider the scalar box, Fig.\ \ref{fig:box} describing a pair production process
in ``deep-inelastic scattering" kinematics,
\bea
p+q \rightarrow K_1+K_2\, ,
\eea
where  incoming line $p$ is massless,  two outgoing lines are massive,
\bea
p^2&=&0\, ,
\nn\\
q^2 &<& 0\, ,
\nn\\
K_1^2 &=& K_2^2 = M^2,
\eea
and where the process is initiated by a space-like momentum transfer, $q$.    In
the notation of Eq.\ (\ref{eq:D-explicit}) and Fig.\ \ref{fig:box}, we have we have four line momenta, $l-p_i$, with
\bea
p_1 &=& 0\, ,
\nn\\
p_2 &=& p\, ,
\nn\\
p_3 &=& p+q\, ,
\nn\\
p_4 &=& K_1\, .
\eea
We assign a mass $M$ to the propagator carrying momentum $l-K_1$,
while other propagators are taken as massless,
\bea
I_{4,1}(\{p_i\},M) = -i\; \int \frac{d^4l}{(2\pi)^4}\, \frac{1}{l^2+i\ep}\,  \frac{1}{(l-p)^2+i\ep}\
\frac{1}{(l-p-q)^2+i\ep}\ \frac{1}{(l-K_1)^2-M^2+i\ep}\, .
\nn\\
\eea
In Minkowski space and with the momenta chosen as above, this integral has a double-logarithmic infrared behavior
when the loop momentum $l$ becomes proportional to $p$ (collinear singularity) with vanishing energy (soft singularity),
and no other sources of double logarithms.   Without fully evaluating the diagram, Fig.\ \ref{fig:box},
let us see how a double-logarithmic behavior emerges in the $(2,2)$ integral.

The term that has double-logarithmic behavior in Eq.\ (\ref{eq:alt-2-N-4}) for this diagram in $(2,2)$ signature is the choice 
$k_i= l$, $k_j= l-p$, that is, the term with the mass shell
poles of the two lines that become parallel.   To be definite, we label $k_{\beta_1}= l-p-q$, $k_{\beta_2}=l-K_1$.
With the routing of momenta shown in the figure, $B_i=0$ and $p_i=0$, so that the relevant term in (\ref{eq:alt-2-N-4}) is
\bea
I_{4,1}^{(l,l-p)} &=& \frac{1}{4(2\pi)^2}\ 
\int_{0}^{l_1^+}  dl^+\  
\int d l^\tp \theta\left(  \{ l, p \} \right)\   \{ l, -p \}  
\nn\\
&\ & \hspace{5mm} \times 
\frac{1}{  - B_{l-p} \{l,l-K_1 \}
 + B_{l-K_1} \{l, l-p \}  + i\epsilon \{ p, K_1 \}  }
 \nn\\
&\ & \hspace{5mm} \times 
\frac{1}{  - B_{l-p} \{l,l+q \}
 + B_{l-q-p} \{l, q-p \}  + i\epsilon \{ p, p+q \}  } \, ,
 \label{eq:alt-2-4-2}
\eea
where we have replaced indices $d$ on the $B_d$ by
the corresponding momenta, $k_d$.
The coefficients of the $B_{k_d}$  are given by
\bea
\{ l, p_i-p_j\} &=& \{l, -p\}  \nn\\ 
&=&  p^+ l^\tp\ -\ l^+ p^\tp \
\nn\\
&=& l^+ p^+ \left( r_l\ -\ r_p \right)\, ,
\label{eq:q-Pij}
\eea
where we have have used the notation of 
Eq.\ (\ref{eq:mudefs}) for $r_l$ and $r_p$.
This antisymmetric combination vanishes both when 
loop momentum $l$ is proportional to the massless 
momentum $p$, so that $r_l=r_p$, and  when $l^+$
vanishes.   These are the collinear and
soft  limits from Minkowski analysis, and the 
limits for $l^+$ and $l^\tp$ are just at these points.
The numerator factor vanishes linearly in both the
collinear and soft limits, but the denominators
with momenta $l-K_1$ and $l-p -q$  behave as
\bea
- \ B_{l-p} \{l,-K_1 \}\
 + \ B_{l-K_1} \{l, -p\} &=& -\ u_1\, (l^+)^2\, (r_l-r_p) + \cdots
 \nn\\
 - \ B_{l-p} \{l, -p-q \}\
 + \ B_{l,-p-q} \{l,l-p\} &=& s\, p^+ l^+(r_l-r_p) + \cdots\ ,
 \eea
 respectively, with $s\equiv (p+q)^2$ and $u_1\equiv 2p\cdot K_1$,
 where neglected terms are higher order in $l^+$ and/or $r_l-r_p$.  
 In deriving these results, we have used that
 $p^2=0$ implies  $p^\tp/p^+ =-p^-/p^\tm$.  
 Now changing variables from $l^\tp$ to $r_l$, we find near 
 the end-points a double-logarithmic integral,
 \bea
 I_{4,1}^{(l,l-p)} &=& -\ \frac{1}{4(2\pi)^2}\  \frac{1}{ u_1\, s}\ \int_{0} \frac{d l^+}{l^+}\  
\int_{r_{p}}  \frac{d r_l}{r_l - r_p}\, .
\label{eq:I-4-2-DL}
\eea
It is straightforward to check that no other term in the sum over poles
has an end-point singularity at $r_l=r_p$, and hence a collinear singularity.

We can compare the result (\ref{eq:I-4-2-DL}) to the double-logarithmic integral
in Minkowski signature, which appears by taking the energy pole
at $l^0 =\sqrt{|\vec l|^2}$ in Fig.\ \ref{fig:box}.   In that case,
in the limit that $\cos\theta_{p l} \rightarrow 1$,  where
$\theta_{pl}$ is the angle between $\vec l$ and $\vec p$,
we find 
\bea
I_{DL} =-\  \frac{1}{4(2\pi)^2}\  \frac{1}{ u_1\, s}\ \int_{0} \frac{d |\vec l|}{|\vec l |}\  
\int^1  \frac{d \cos\theta_{pl}}{1 - \cos\theta_{p l}}\, ,
\eea
with the same double-logarithmic behavior as (\ref{eq:I-4-2-DL}) up to
a change of variables.

In the above calculation, we have not discussed regulation of  infrared-divergent integrals.
The simplest regulation for the example above is to take $p_1^2<0$, but with gauge theories
in mind it is natural to ask whether dimensional regularization is possible.   
Although our approach to $(2,2)$ signature is closely linked to four dimensions, there
is in fact nothing to keep us from dimensionally regulating.    The interpretation is
particularly straightforward for infrared regulation, which requires $\vep = 2 - D/2<0$,
with $D$ the number of dimensions,  taken greater than four.
We thus imagine adding $-2\vep$ dimensions to the four dimensions spanned
by our coordinates $l^\pm$ and $l^{\tilde \pm}$.

While a full discussion of dimensional regularization 
for multi loop diagrams would take an extensive
analysis, we will content ourselves here with the observation that if we 
label the momenta of the extra dimensions as $l_\perp$, and keep the
external momenta in four dimensions, all of the analysis leading to our
one loop result, Eq.\ (\ref{eq:alt-2-N-4}), for example, is unchanged.
The effect of dimensional regularization
is simply to add a term $-l_\perp^2$ to every squared mass term in the denominators of (\ref{eq:alt-2-N-4}),
$B_\alpha \rightarrow B_\alpha - l_\perp^2$ in Eq.\ (\ref{eq:AB-coeff-def}),
and to introduce an overall integration over the ``extra" dimensions of the form
\bea
\frac{2\pi^\vep}{\Gamma(\vep)}\ \int_0^\infty dl_\perp\, l_\perp^{-2\vep - 1}\, ,
\eea
acting on the modified integrand,
where the prefactor represents the angular volume.   In the limit
$\vep\rightarrow 0$, the zero of the angular integration is balanced
by the (infrared) pole from the radial integral.    For infrared finite 
integrands, the net result is unity for $\vep=0$, but for divergent 
integrals as in Eq.\ (\ref{eq:I-4-2-DL}), the result is infrared regulated
after the $l_\perp$ integration.

\section{Summary and Conclusions}

We have studied scalar perturbation theory in $(2,2)$ signature, and have
identified a natural analytic continuation from Minkowski signature, which
crosses no singularities and can be used to define diagrams with arbitrary
external momenta.    The resulting integrals  have
a standard ``$i\ep$" prescription for the definition of contours in
the presence of propagator singularities.  This enables us to appeal
to standard Landau analysis to identify pinches of momentum
integrals, and singularities in external momenta.   The singularities
in $(2,2)$ are in general quite different than those in $(1,3)$ signature.
  An exception is when external momenta are restricted to
a plane in Minkowski space; in this case the contour rotation to $(2,2)$
signature does not change the integral.

For diagrams that are fully ultraviolet finite (in all subdiagrams), we can
introduce  two sets of light cone variables, all four of which are linear in all denominators.
We have derived a general expression for such an
$L$-loop $N$-line integral as the sum of $2L$-dimensional integrals using
$(2,2)$ integration.
Whether these expressions can be of use in the practical
evaluation of higher-loop scalar integrals is a subject for further investigation.   

\begin{acknowledgments}
We thank S.\ Caron-Huot and E.\ Witten for useful conversations.
This work was supported by the
National Science Foundation,  grant  PHY-0969739.   GS thanks the
European Centre for Theoretical Studies in Nuclear Physics and Related
Areas for hospitality during the workshop: 
``Scattering Ampliutdes: from QCD to Maximally Supersymmetric Yang-Mills Theory and Back",
which played a role in the development of this project.
\end{acknowledgments}

\end{document}